\documentclass{article}
\usepackage{spconf,amsmath,amssymb,graphicx,hyperref}

\title{Multiscale Gaussian-Mixture Modeling for HMM Post-Processing in Selective Auditory Attention Decoding}

\name{Tianyi Li$^{1}$, Simon Geirnaert$^{1,2}$, Bert De Smedt$^{3}$, and Alexander Bertrand$^{1}$\thanks{This work was supported by Internal Funds KU Leuven (projects IDN/23/006, C14/25/108, and C3/25/107), FWO projects G081722N and G026026N, and the FWO Junior Postdoctoral Fellowship for Fundamental Research awarded to S. Geirnaert (No. 1242524N). The authors thank Davina Van den Broek and Elien Bellon for their collaboration in the broader research project.}}

\address{
$^{1}$KU Leuven, Department of Electrical Engineering (ESAT), Leuven, Belgium\\
$^{2}$KU Leuven, Department of Neurosciences, Research Group ExpORL, Leuven, Belgium\\
$^{3}$KU Leuven, Faculty of Psychology and Educational Sciences, Leuven, Belgium.
}

\begin{document}
\ninept

\maketitle

\begin{abstract}

Selective auditory attention decoding (sAAD) infers from electroencephalography (EEG) which speaker a listener attends to in multi-speaker scenarios. A widely used approach reconstructs the attended speech envelope from EEG, correlates it with candidate envelopes, and selects the speaker with the highest Pearson correlation. However, correlations in each decision window have high intrinsic variance, making per-window decisions unreliable, especially for short windows. Recent work introduced a hidden Markov model (HMM) that integrates correlation evidence over time and requires state-dependent emission distributions. Since labeled attention states are typically unavailable in practice, these distributions can be inferred from unlabeled observations by fitting a Gaussian mixture model (GMM). However, when the underlying Gaussians are close, unsupervised fitting becomes less reliable with limited samples. We therefore propose a multiscale GMM whose parameters are jointly estimated across multiple window lengths. The model exploits the window-length dependence of Fisher-transformed correlation estimates, whose means are approximately stable while their variances decrease with the number of samples per window. Experiments show that the multiscale GMM estimates emission parameters more accurately than single-scale fitting, particularly at short windows. Its HMM-post-processed accuracy advantage is largest for short recordings and narrows as more data become available, motivating multiscale fitting in data-constrained settings.

\end{abstract}


\section{Introduction}
\label{sec:intro}

Real-world listening often involves multiple concurrent speakers, where listeners need to focus on a speaker of interest while ignoring competing speakers. Neural tracking studies have shown that attended speech is represented more strongly than unattended speech in cortical responses~\cite{mesgarani2012selective,power2012what}. Identifying the attended speaker from electroencephalography (EEG), known as selective auditory attention decoding (sAAD), has been widely studied in recent years, with potential applications in neuro-steered hearing devices~\cite{mirkovic2016target}.

A common sAAD strategy is stimulus reconstruction~\cite{o2015attentional,biesmans2016auditory,de2018decoding,crosse2016multivariate}, where a neural decoder is trained to reconstruct the attended speech envelope from EEG, and the candidate speaker whose envelope yields the highest Pearson correlation with this reconstruction is identified to be the attended speaker. sAAD has also been studied under more realistic acoustic conditions and with more wearable EEG configurations~\cite{mirkovic2016target,fuglsang2017noise}. However, per-window correlations remain noisy, especially for short decision windows of one or a few seconds~\cite{heintz2025post,geirnaert2025performance}.

To improve sAAD performance, recent work has proposed a post-processing algorithm based on a hidden Markov model (HMM), which integrates correlation evidence over time~\cite{heintz2025post}. In this framework, the attention state, indicating which speaker is attended, evolves according to a transition probability that reflects the assumption that listeners switch attention relatively infrequently between adjacent windows. The observed correlation values are then modeled generatively through state-dependent emission distributions, which describe how likely different correlation patterns are under each attention state. The resulting predictions have been shown to be more accurate than raw decoder decisions without compromising switch detection time~\cite{heintz2025post}.

While the transition probability can be set as a hyperparameter, the state-dependent emission distributions must be estimated from data~\cite{heintz2025post}. A supervised estimate would require windows with known attention states, but such labels are typically unavailable in practice. A natural unsupervised alternative, proposed by Heintz et al.~\cite{heintz2025post}, is a Gaussian mixture model (GMM): for each participant, the Fisher-transformed correlation is modeled by a two-component GMM, with the higher-mean component corresponding to the distribution of the correlation with the attended speaker and the lower-mean component corresponding to the distribution of the correlation with the unattended speaker. This assignment follows from the observation that EEG responses generally track attended speech more strongly than unattended speech~\cite{mesgarani2012selective,power2012what,o2015attentional}.

HMM-based sAAD is particularly attractive at short operating windows because
they preserve the temporal resolution needed to detect attention switches \cite{heintz2025post}.
Yet short-window correlations are noisy, making their emission distributions
difficult to estimate reliably with a single-scale GMM. We address this problem by using longer-window correlations to assist the
estimation of short-window HMM emissions. Correlations at different window
lengths reflect the same underlying attended and unattended neural responses, but with different estimation noise variances.
After Fisher transformation, their means are approximately invariant across
window lengths, whereas their variances scale approximately inversely with the
number of samples~\cite{geirnaert2025performance, fisher1921probable}.
Based on these properties, the proposed \emph{multiscale GMM} jointly estimates
the emission parameters from observations at all window lengths under this
scaling law.

The proposed method is particularly useful in two settings. First, for
short-window operation, the multiscale GMM incorporates longer-window
correlation observations into the estimation of short-window emissions.
Second, with limited available data, its cross-scale constraints regularize
GMM estimation. We evaluate these settings by measuring emission-parameter
recovery across window lengths and HMM-post-processed classification accuracy
at different available recording durations.


\section{Multiscale GMM}
\label{sec:method}

This section reviews HMM post-processing and introduces the multiscale GMM used to estimate its state-dependent emission distributions. We start from a standard stimulus-reconstruction sAAD pipeline with 2 speakers. A backward decoder reconstructs the attended speech envelope from EEG and can be trained in either a supervised~\cite{o2015attentional,biesmans2016auditory} or unsupervised~\cite{geirnaert2021unsupervised} fashion. Within each decision window of length $L$, we compute two Fisher-transformed Pearson correlations $\mathbf{z}=(z_1,z_2)$ (hereafter Fisher-z correlations),
\begin{equation}
z_i = \operatorname{atanh}\bigl(\operatorname{corr}(\hat{y}, y_i)\bigr),
\quad i \in \{1,2\},
\label{eq:zs}
\end{equation}
where $\hat{y}$ is the reconstructed speech envelope (decoded from EEG recordings) and $y_i$ is the
envelope of candidate speaker $i$. If speaker 1 is the attended speaker, then we expect $\operatorname{corr}(\hat{y}, y_1)>\operatorname{corr}(\hat{y}, y_2)$ (hence $z_1>z_2$ due to monotonicity of the Fisher transform) and vice versa. The Fisher transformation makes the correlation distribution approximately
Gaussian~\cite{fisher1921probable}.

For a sequence of $T$ decision windows, let $\mathbf{z}_{1:T}=\{\mathbf{z}_t\}_{t=1}^{T}$ denote the Fisher-z correlation observations, and let $s_t\in\{1,2\}$ denote the attended speaker at window $t$. The HMM postprocessor defined in \cite{heintz2025post} combines a transition model $p(s_t\mid s_{t-1})$ with an emission model $p(\mathbf{z}_t\mid s_t)$. We use forward-backward inference to compute $p(s_t\mid \mathbf{z}_{1:T})$ and select the state with the largest posterior probability.

Following~\cite{heintz2025post}, we assume conditional independence between
$z_1$ and $z_2$ given the attention state $s$. The two correlation streams follow
the attended density $p(z\mid\mathrm{att})$ and unattended density
$p(z\mid\mathrm{unatt})$ according to the attention state. The HMM emission
density then factorizes as
\begin{equation}
p(z_1,z_2 \mid s) =
\begin{cases}
p(z_1 \mid \mathrm{att})\,p(z_2 \mid \mathrm{unatt}), & \text{if } s=1,\\
p(z_2 \mid \mathrm{att})\,p(z_1 \mid \mathrm{unatt}), & \text{if } s=2.
\end{cases}
\label{eq:emission-factorization}
\end{equation}
The attended and unattended densities are modeled as Gaussian components with
distinct means $\mu_{\mathrm{att}}$ and $\mu_{\mathrm{unatt}}$ and a common\footnote{For simplicity, and as empirically shown in \cite{lopez2025unsupervised}, both components share the same $\sigma_L$, which will also be theoretically and empirically motivated further on (see equation \eqref{eq:hotelling} and Fig. \ref{fig:empirical-statistics}, respectively).} scale-dependent variance $\sigma_L^2$, where $L$ is the decision window length over which the correlation is computed.
\begin{equation}
\begin{aligned}
p(z|\mathrm{att})
&= \mathcal{N}(z;\mu_{\mathrm{att}},\sigma_L^2),\\
p(z|\mathrm{unatt})
&= \mathcal{N}(z;\mu_{\mathrm{unatt}},\sigma_L^2).
\end{aligned}
\label{eq:conditional-gaussians}
\end{equation}
The attended mean is assumed to be larger than the unattended mean, because
the EEG reconstruction is expected to correlate more strongly with attended
speech~\cite{o2015attentional}.

\subsection{Distributional properties of Fisher-z correlations}
\label{sec:law}

A conventional single-scale GMM estimates the emission model independently at
the operating window length $L$. In this work, we instead exploit the
approximate distribution of Fisher-z correlations and its dependence on $L$.
For a window of length $L$, let $N_L=Lf_s$ denote the number of samples
used to compute the correlation, where $f_s$ is the sampling rate. For a Pearson correlation coefficient $r$ with
underlying true correlation $\rho$, the standard Fisher-Hotelling
approximation models the Fisher-z statistic $z=\operatorname{atanh}(r)$ as
approximately Gaussian with mean and variance given by:
\begin{equation}
\mathbb{E}[z \mid \rho,L]
\approx
\operatorname{atanh}(\rho)
+
\frac{\rho}{2(N_L-1)},
\quad
\operatorname{Var}(z \mid \rho,L)
\approx
\frac{1}{N_L-1}.
\label{eq:hotelling}
\end{equation}
This type of correlation-distribution modeling has been used to characterize
AAD performance across window lengths~\cite{geirnaert2025performance}. Related
scale-dependent behavior of correlation statistics was examined by Lopez-Gordo
et al.~\cite{lopez2025unsupervised} for unsupervised AAD accuracy estimation.
Here, we apply this idea to the Fisher-z correlations used as HMM observations.

\textbf{Approximate mean stability.}\;
The dominant term in~\eqref{eq:hotelling} is $\operatorname{atanh}(\rho)$,
which depends on the underlying attended or unattended correlation regardless of the window length. The first-order Hotelling correction introduces only a weak
scale dependence,
$\operatorname{atanh}(\rho)+\rho/[2(N_L-1)]$. For the sampling rate and window
lengths considered here, this term is small compared with the dominant
Fisher-z mean. A corrected-mean sensitivity analysis also had negligible effect.
We therefore use the scale-invariant mean model:
\begin{equation}
\forall\:L:
\mu_{\mathrm{att}}(L)\approx\mu_{\mathrm{att}}, \quad
\mu_{\mathrm{unatt}}(L)\approx\mu_{\mathrm{unatt}}.
\label{eq:mu-invariance}
\end{equation}

\textbf{Variance scaling.}\;
Following the inverse-sample-size dependence in~\eqref{eq:hotelling}, the GMM
component variance is modeled as
\begin{equation}
\sigma_L^2 = \frac{C}{N_L-1}.
\label{eq:sigma-law}
\end{equation}
Under the ideal Hotelling approximation, $C=1$. Here, estimating $C$ as an
unknown parameter allows the model to capture the empirical variance level
while preserving the expected dependence on window length.

Together, \eqref{eq:mu-invariance} and~\eqref{eq:sigma-law} imply that
correlations measured at different window lengths follow a parametric family
governed by a small set of parameters shared across window lengths. These two
properties are empirically examined in Section~\ref{sec:result}.

\subsection{Proposed multiscale GMM algorithm}
\label{sec:multiscale-gmm}

Let $\mathcal{L}=\{L_1,\ldots,L_K\}$ be the set of window lengths used in the multiscale GMM fitting. For each window length $L\in\mathcal{L}$, let
$\{z_t^{(L)}\}_{t=1}^{T_L}$ contain the Fisher-z observations pooled from both
candidate speakers. Here, $T_L$ counts the pooled scalar observations, which depends on the window length $L$ since windows are not overlapping. The
multiscale GMM parameters are
$\theta=\{\mu_{\mathrm{att}},\mu_{\mathrm{unatt}},C,\pi\}$,
where $\mu_{\mathrm{att}}$ and $\mu_{\mathrm{unatt}}$ are the scale-invariant
component means, $C$ is the variance scale parameter, and $\pi$ is the mixture
weight of the GMM. The mixture density at scale $L$ is
\begin{equation}
\begin{split}
p(z \mid \theta,L) ={} & \pi\,\mathcal{N}(z \mid \mu_{\mathrm{att}}, \sigma_L^2) \\
& + (1-\pi)\,\mathcal{N}(z \mid \mu_{\mathrm{unatt}}, \sigma_L^2),
\end{split}
\end{equation}
with $\sigma_L^2=C/(N_L-1)$. The parameters are estimated by maximizing the composite log-likelihood pooled across scales,
\begin{equation}
\hat{\theta}=\arg\max_{\theta}\sum_{L\in\mathcal{L}}\sum_{t=1}^{T_L}
\log p(z_t^{(L)} \mid \theta,L).
\label{eq:objective}
\end{equation}

We solve \eqref{eq:objective} using expectation-maximization (EM). The
component means are initialized at the $0.33$ and $0.67$ empirical quantiles
of the pooled observations, $C$ is initialized such that the average of \eqref{eq:sigma-law} across scales corresponds to the empirical average
, and
$\pi^{(0)}=1/2$. Although the present data are balanced, we keep $\pi$ learnable for additional flexibility in the unsupervised fit. The lower- and higher-mean components are assigned to the
unattended and attended states, respectively.

The E-step computes the responsibility of the attended component,
\begin{equation}
\gamma_t^{(L)} =
\frac{\pi\,\mathcal{N}(z_t^{(L)} \mid \mu_{\mathrm{att}}, \sigma_L^2)}
{\pi\,\mathcal{N}(z_t^{(L)} \mid \mu_{\mathrm{att}}, \sigma_L^2)
+ (1-\pi)\,\mathcal{N}(z_t^{(L)} \mid \mu_{\mathrm{unatt}}, \sigma_L^2)}.
\end{equation}
Since $N_L-1$ is proportional to the precision at scale $L$, the M-step updates
for the component means and variance scale are
\begin{align}
\mu_{\mathrm{att}} &=
\frac{\sum_{L,t} (N_L-1)\,\gamma_t^{(L)}\,z_t^{(L)}}
{\sum_{L,t} (N_L-1)\,\gamma_t^{(L)}},\\
\mu_{\mathrm{unatt}} &=
\frac{\sum_{L,t} (N_L-1)\,(1-\gamma_t^{(L)})\,z_t^{(L)}}
{\sum_{L,t} (N_L-1)\,(1-\gamma_t^{(L)})},\\
C &= \frac{1}{T_{\mathrm{tot}}}\sum_{L,t}(N_L-1)
\Big[\gamma_t^{(L)}(z_t^{(L)}-\mu_{\mathrm{att}})^2 \notag\\
&\quad +(1-\gamma_t^{(L)})(z_t^{(L)}-\mu_{\mathrm{unatt}})^2\Big],\\
\sigma_L^2 &= \frac{C}{N_L-1}.
\end{align}
The mixture weight is updated as
$\pi=T_{\mathrm{tot}}^{-1}\sum_{L,t}\gamma_t^{(L)}$.
After each M-step, the components are reordered if necessary to maintain
$\mu_{\mathrm{att}}>\mu_{\mathrm{unatt}}$, and the iterations are stopped when
the relative log-likelihood increase falls below a preset tolerance or the
maximum number of iterations is reached. The multiscale GMM estimates four
parameters from data pooled across all $K$ scales, instead of
$4K$ parameters for independent single-scale GMMs. At
inference time, the fitted densities are inserted into
\eqref{eq:emission-factorization} to construct the two HMM-state emissions.
\section{Experimental Setup}
\label{sec:setup}

All experiments use a supervised leave-one-participant-out (LOSO) decoder
trained on the other 15 participants. GMM fitting and HMM post-processing are
performed separately for each held-out participant, with no parameter sharing
across participants. Held-out labels are used only for evaluation, not for GMM
fitting.

\subsection{Dataset}
We evaluate the method on the publicly available KU Leuven sAAD
dataset~\cite{biesmans2016auditory,das2019auditory}, containing $64$-channel
EEG from $16$ normal-hearing participants listening to two competing Dutch
speakers. Each participant completed $20$ trials with one attended speaker.
After truncation to whole minutes, eight trials are $6$~min and twelve are
$2$~min, yielding $72$~min. In each LOSO fold, all 20 trials from each of the
15 training participants are used to train the decoder. All 20 trials from the
held-out participant are used for GMM fitting and HMM post-processing.

\subsection{Preprocessing}
We follow the public KUL preprocessing pipeline~\cite{biesmans2016auditory}.
Speech envelopes are extracted from the original speech signals using a gammatone
filterbank and power-law compression with exponent $0.6$. Both EEG and
envelopes are bandpass filtered to $1$-$9$~Hz, downsampled to
$f_s=32$~Hz, and the EEG is re-referenced to Cz.

\subsection{Decoder training}
Following~\cite{biesmans2016auditory}, we use a spatio-temporal least-squares backward decoder to reconstruct the attended
speech envelope from EEG. The decoder uses $N_c=64$ channels and
$J=17$ post-stimulus lags, corresponding to $0$-$500$~ms at
$f_s=32$~Hz~\cite{vanthornhout2019effect}. For each LOSO fold, the normal
equations are pooled over data from the 15 training participants and
solved to obtain one subject-independent decoder. It is applied to the held-out participant to obtain
the reconstructed envelope, which is correlated with each candidate envelope
and Fisher-transformed as in~\eqref{eq:zs}. 
\subsection{Multiscale GMM and HMM inference}
On the evaluation trials, we compute Fisher-z correlations at $K=8$ window
lengths, $L\in\mathcal{L}=\{1,2,3,5,6,10,15,30\}$~s.
These window lengths cover short, intermediate, and long windows while
allowing each retained trial to be divided into complete windows at every
scale.
The multiscale GMM is fitted jointly over all $K$ window lengths.

We compare the proposed multiscale GMM with the single-scale GMM used in \cite{heintz2025post} that fits the
emission model at window length $L$, using the same EM
initialization and update rules. For each target $L$, the evaluation
trials are concatenated trial by trial to form one sequence per participant.
Forward-backward inference uses a symmetric transition probability
$p_{\mathrm{switch}}(L)=0.01L$.

Emission-parameter recovery is reported at all eight scales using the complete
72-min fitting set. For HMM post-processing, we focus on short operating
windows at $L^*=1$ and $2$~s as suggested by \cite{heintz2025post}, while
longer-window observations serve as auxiliary data in the multiscale fit. To
assess performance as a function of available data, we vary the recording
duration from 6 to 72~min in 6-min steps. At each duration, accuracy is averaged
within participant over up to ten data-subset realizations. Trials within each
realization are ordered so that the attended speaker alternates at every trial
boundary. Each realization is used for both GMM fitting and HMM post-processing
and is identical for both methods.

\subsection{Evaluation metrics}
\label{sec:metrics}

Emission-parameter estimation is evaluated against label-based empirical
references. For each participant and window length, observations from both
candidate streams are pooled across both speakers and split according to whether the corresponding
speaker was attended or unattended. Their sample means and pooled standard
deviation are used as supervised ``ground-truth'' references for the Gaussian
parameters.

Specifically, we evaluate
$a\in\{\mu_{\mathrm{att}},\mu_{\mathrm{unatt}},\sigma^*,\Delta\mu\}$,
where $\Delta\mu=\mu_{\mathrm{att}}-\mu_{\mathrm{unatt}}$ and
$\sigma^*=\sqrt{C/(N_{L^*}-1)}$, with $L^*$ denoting the target
correlation window length used by the HMM. Let
$a^{\mathrm{fit}}$ and $a^{\mathrm{ref}}$ denote the fitted and empirical values,
respectively. Parameter recovery is quantified by the normalized relative error
$E_a=|a^{\mathrm{fit}}-a^{\mathrm{ref}}|/|a^{\mathrm{ref}}|$.

HMM-post-processed classification accuracy is computed within each trial as
the percentage of decision windows for which the HMM's maximum-posterior state
matches the true attended speaker, and then averaged across trials for each
participant~\cite{heintz2025post}. Paired comparisons at the 6- and 72-min recording-duration endpoints
use one-sided Wilcoxon signed-rank tests for higher multiscale accuracy,
with Holm correction across the four comparisons arising from the two
target window lengths ($L^*=1$ and $2$~s) and the two recording-duration
endpoints. Emission-parameter recovery and intermediate data-amount trends are otherwise
summarized using the mean and standard error of the mean (SEM) across
participants.

\section{Results and discussion}
\label{sec:result}

\begin{figure}[t]
\centering
\includegraphics[width=\linewidth]{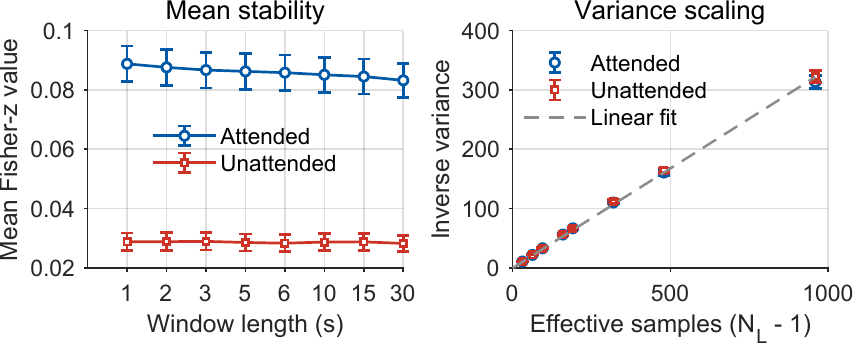}
\caption{Empirical attended and unattended Fisher-z means (left) and inverse
variance versus $N_L-1$ (right), averaged across participants. Error bars
denote mean $\pm$ SEM; the dashed line is a linear fit through the origin.}
\label{fig:empirical-statistics}
\end{figure}

Figure~\ref{fig:empirical-statistics} examines the two multiscale assumptions.
For this diagnostic only, the true labels are used to pool observations into
attended and unattended groups. Their means remain separated and vary only
weakly with window length. Meanwhile, inverse variance is approximately linear
in $N_L-1$. Long-window observations can therefore inform the component means
at short operating windows, while the variance law maps their precision across
scales.

\begin{figure}[t]
\centering
\includegraphics[width=\linewidth]{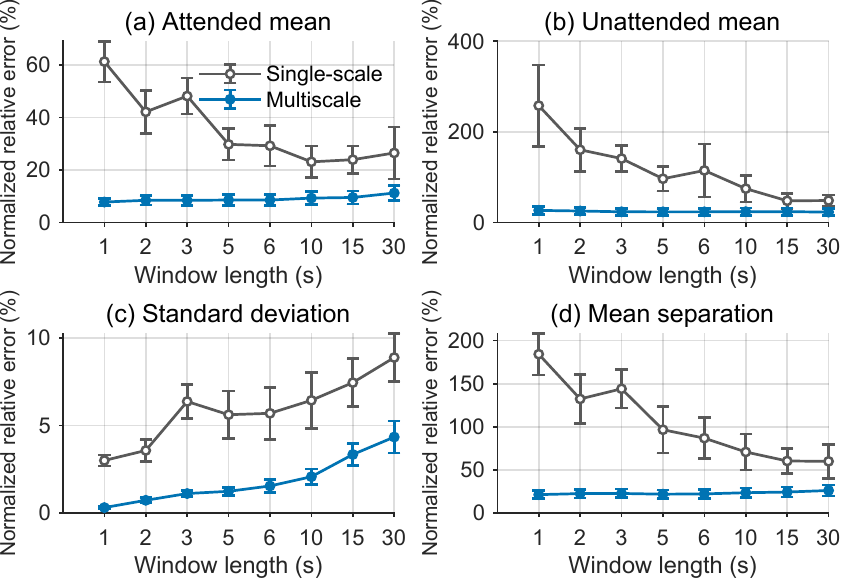}
\caption{Normalized relative error across window lengths for the attended
mean, unattended mean, standard deviation, and mean separation
$\Delta\mu=\mu_{\mathrm{att}}-\mu_{\mathrm{unatt}}$. Markers and error bars
denote mean $\pm$ SEM across participants.}
\label{fig:gmm-error}
\end{figure}

As shown in
Fig.~\ref{fig:gmm-error}, multiscale fitting yields lower mean errors than
single-scale fitting for all four evaluated quantities at both $L^*=1$ and $L^*=2$~s,
with larger reductions at 1~s. At longer windows, lower correlation variance
separates attended and unattended samples more clearly, making the
single-scale GMM easier to fit. Thus, long windows are most valuable as
auxiliary data for stabilizing noisy short-window HMM emissions.

We next evaluate HMM post-processing under the default full-data setting.
The mean classification accuracies for
single-scale and multiscale fitting were 76.17\% and 77.98\% at 1~s,
respectively, corresponding to a difference of 1.82~percentage points (pp)
(Holm-corrected $p=0.0077$).
At 2~s, the corresponding accuracies were 77.24\% and 78.32\%, with a
difference of 1.08~pp (Holm-corrected $p=0.0232$).

\begin{figure}[!h]
\centering
\includegraphics[width=0.8\linewidth]{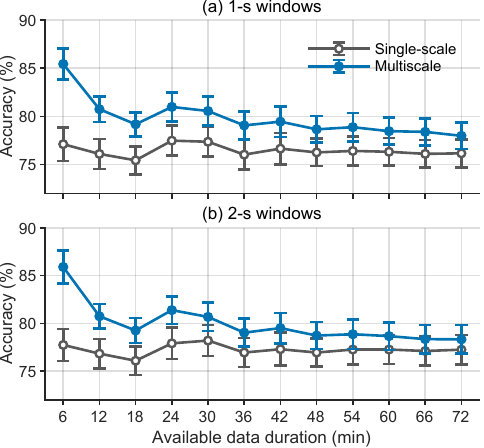}
\caption{HMM-post-processed classification accuracy at 1- and 2-s windows
versus available recording duration (mean $\pm$ SEM across participants,
averaged over data-subset realizations).}

\label{fig:data-efficiency}
\end{figure}

Figure~\ref{fig:data-efficiency} shows how classification accuracy changes
with available recording duration. With only 6~min of data, the mean
multiscale-minus-single-scale accuracy
differences are 8.32~pp at 1~s and 8.16~pp at 2~s (both Holm-corrected
$p<0.001$). As the recording duration increases, the two curves gradually
approach each other. Shorter subsets contain fewer trial boundaries and
therefore fewer attention-state changes, which affects the absolute accuracy
levels.
The larger paired differences at short durations highlight the value of
multiscale fitting when only a short recording is available.

\section{Conclusion}
\label{sec:conclusion}

We have proposed a multiscale GMM that jointly estimates shared emission parameters from correlation observations at multiple window lengths. This allows the more reliable statistics at longer windows to improve the estimation of the HMM emission probabilities based on noisy short-window correlations. Experiments showed more accurate short-window emission-parameter recovery than single-scale fitting. The data-amount analysis further showed larger mean classification-accuracy gains for shorter available recordings, highlighting the value of multiscale estimation in data-constrained settings.

\clearpage

\bibliographystyle{IEEEbib}
\bibliography{ref}

\end{document}